\documentclass[floatfix,prl,twocolumn]{revtex4}
\usepackage{textcomp}
\usepackage{graphicx}
\usepackage{amsmath}
\usepackage{amssymb}
\usepackage{color}

\begin{document}

\title{Bubble Raft Model for a Paraboloidal Crystal}

\author{Mark J. Bowick}
\author{Luca Giomi}
\author{Homin Shin}
\author{Creighton K. Thomas}

\affiliation{Department of Physics,
Syracuse University,
Syracuse New York,
13244-1130}

\begin{abstract}
We investigate crystalline order on a two-dimensional paraboloid of
revolution by assembling a single layer of millimeter-sized soap
bubbles on the surface of a rotating liquid, thus extending the
classic work of Bragg and Nye on planar soap bubble rafts.
Topological constraints require crystalline configurations to
contain a certain minimum number of topological defects such as
disclinations or grain boundary scars whose structure is analyzed as
a function of the aspect ratio of the paraboloid. We find the defect
structure to agree with theoretical predictions and propose a
mechanism for scar nucleation in the presence of large Gaussian
curvature.
\end{abstract}

\maketitle

Soft materials such as amphiphilic membranes, diblock copolymers and
colloidal emulsions can form ordered structures with a wide range of
complex geometries and topologies. Macroscopic models of ordered
systems of this type are desirable for direct visualization and table
top demonstrations, and they can be used as control checks of
theoretical predictions.  In this paper, we discuss the fabrication of
a paraboloidal soap bubble raft which realizes a two-dimensional
crystalline monolayer with both variable Gaussian curvature and a
boundary.

Some 60 years ago Bragg and Nye used bubble rafts to model metallic
crystalline structures \cite{BraggNye:1947}. A carefully made
assemblage of bubbles, floating on the surface of a soap solution
and held together by capillary forces, forms an excellent
two-dimensional replica of a crystalline solid, in which the regular
triangular arrangement of bubbles is analogous to the close packed
structure of atoms in a metal \cite{Kittel:1966}.
Feynman considered this technique to be important enough that the 
famous Feynman lectures in physics include a reproduction
of the original Bragg-Nye paper in its entirety
\cite{Feynman:1963}.
Bubble rafts can be made easily and inexpensively, equilibrate quickly,
exhibit topological defects such as disclinations, dislocations and
grain boundaries, and provide vivid images of the
structure of defects. Bubble raft models
have been used to study two-dimensional polycrystalline and
amorphous arrays \cite{SimpsonHodkinson:1972},
nanoindentation of an initially defect-free crystal
\cite{GouldstoneEtAl:2001}, and the dynamic behavior of crystals
under shear \cite{WangKrishanDennin:2006}.
Beyond these advantages, rotating the soap solution
with bubbles on the surface provides a flexible playground for creating
crystalline order on a nearly perfect paraboloid.

The interplay between order and geometry has been intensively studied in
many systems, including large spherical crystals
\cite{BowickNelsonTravesset:2000},
toroidal hexatics \cite{BowickEtAl:2004}, both crystals and hexatics
draped over a Gaussian bump \cite{VitelliNelson:2004,VitelliLucksNelson:2006},
and paraboloidal crystals \cite{GiomiBowick:2007}. Topological defects are
essential in understanding the crystalline order in a curved
two-dimensional manifold. In some cases (e.g. the sphere) the topology
requires that
a certain minimum number of defects be present in the ground state.
For a two-dimensional Riemannian manifold $M$ with boundary $\partial M$, a
discrete version of the Gauss-Bonnet theorem for any triangulation of $M$ reads
\begin{equation}
Q=\sum_{i\in M} (6-c_{i}) + \sum_{i\in\partial M} (4-c_{i})=6\chi\,,
\end{equation}
where $c_{i}$ is the coordination number of vertex $i$ and $\chi$
is the Euler characteristic. The quantity $q_{i,\,M}=6-c_{i}$ is the
disclination charge for a site $i$ in the interior and measures
the departure from perfect triangular order. The analogous quantity
on the boundary is $q_{i,\,\partial M}=4-c_{i}$. $Q$ thus represents
the total disclination charge. For crystals on the $2-$sphere ($\chi=2$),
$Q =12$, while for crystals on the $2-$torus ($\chi=0$), $Q=0$.
For the disk topology relevant for our experiment, $\chi=1$ and the
total disclination charge $Q=6$. Provided we restrict ourselves to
the energetically preferred minimal $q=\pm 1$ disclinations, we
see that any paraboloidal crystal must have at least six $+1$
disclinations \cite{Diskpapers}.

In the regime of a sufficiently large number of particles, the
isolated disclinations required by the topology are unstable to
grain boundary ``scars'', consisting of  arrays of tightly bound
$5-7$ pairs radiating from an unpaired $+1$ disclination. The
existence of scars, first predicted in the context of spherical
crystallography \cite{BowickNelsonTravesset:2000} and later observed
experimentally in colloidal suspensions on spherical droplets
\cite{BauschEtAl:2003,EinertEtAl:2005}, has become one of the
fundamental signatures of dense geometrically frustrated systems.
The possibility of a coexistence of isolated defects and scars was
also pointed out \cite{GiomiBowick:2007} as a consequence of a
variable Gaussian curvature in both frustrated and unfrustrated
systems.

Calling $z$ the height of the surface above the $xy$ plane, a
paraboloid is straightforwardly parametrized by the function
$z(r) = \frac{h}{R^{2}}\,r^{2}$, where $r$ is the polar distance
on the $xy$ plane, $h$ the height of the paraboloid and $R$ the
maximum radius. In order to provide a position-independent notion
of curvature, it is convenient to introduce the parameter
$\kappa=2h/R^{2}$ representing the normal curvature of the
paraboloid at the origin. For a rotating fluid in a cylindrical
vessel $\kappa=\omega^{2}/g$, where $\omega$ is the angular
velocity of the vessel and $g$ the gravitational acceleration.

\begin{figure*}[t]
\centering
\includegraphics[scale=0.7]{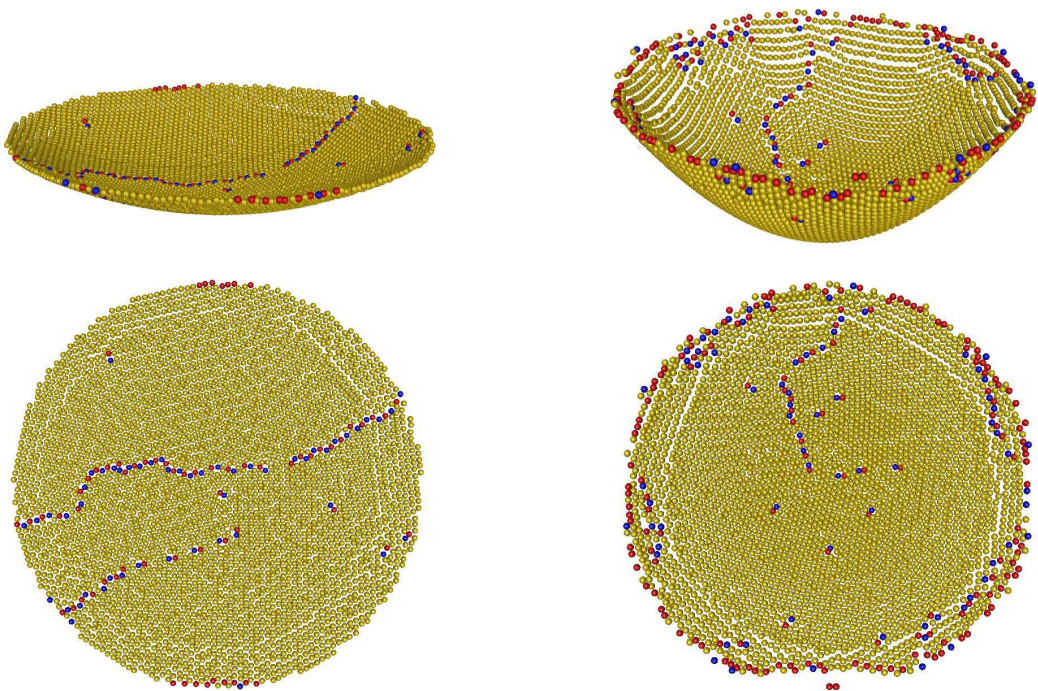}
\caption{\label{fig:raft} (color online)
Lateral and top view of a computer reconstruction
of two paraboloidal rafts with $\kappa_{1} \approx 0.15$ cm$^{-1}$ (left)
and $\kappa_{2} \approx 0.32$ cm$^{-1}$ (right). The number of
bubbles is $N_{1}=3813$ and $N_{2}=3299$ respectively. The color scheme
highlights the $5-$fold (red) and $7-$fold
(blue) disclinations over $6-$fold coordinated bubbles (yellow).}
\end{figure*}

To make the bubble rafts, we pump air through a needle into soapy water.
Because the larger bubble sizes we prefer are most easily made when
the vessel is still, we first make the bubbles and only
later spin the vessel to make the paraboloid (cf.
Bragg and Nye \cite{BraggNye:1947}, who
spun their system in order to generate smaller
bubbles but stopped the spinning to look at the
bubbles on a flat surface).  To image the bubbles, we mount a CCD
digital camera on the top of the vessel, with lighting from
a ring around the (clear) vessel to eliminate glare.  The camera
rotates along with the whole system so that the shutter speed is
unimportant in imaging the bubbles.  We use a second camera to find
the aspect ratio of the paraboloid. We equilibrate the system and
eliminate stacking of bubbles by imposing small perturbations of the
angular frequency to mimic the role of thermal noise.
The vessel has radius $R=5$ cm; the height of paraboloids
varies from $h=0$--$4$ cm. The bubble diameter,
extracted from the Delaunay triangulation of our images, is
$a=0.84(1)$ mm with monodispersity $\Delta a/a \approx 0.003$.
The normal curvature $\kappa$ of the paraboloid at the origin varies
from $0$--$0.32$ cm$^{-1}$. In addition to the flat disk, we
observe two different curvature regimes: small curvature
$\kappa_{1}\approx 0.15$ cm$^{-1}$ and high curvature
$\kappa_{2}\approx 0.32$ cm$^{-1}$.

Figure \ref{fig:raft} shows a computer reconstruction of two bubble
rafts with these $\kappa = \kappa_{1}$, $\kappa_{2}$.
We extract two dimensional coordinates from the
images with
a brightness based particle location algorithm \cite{CrockerGrier:1996}.
Data sets are then processed to correct possible imprecisions and finally
Delaunay triangulated. We choose to exclude from the triangulation the first 
$3$--$4$ bubble rings formed along the boundary of the cylindrical vessel, 
where the sharp concave meniscus due to the surface tension combined with 
the native curvature of the paraboloid was observed to produce a stacking 
of bubbles in a narrow double layer surrounding the perimeter of the vessel.

\begin{figure}[t]
\centering
\includegraphics[width=1\columnwidth]{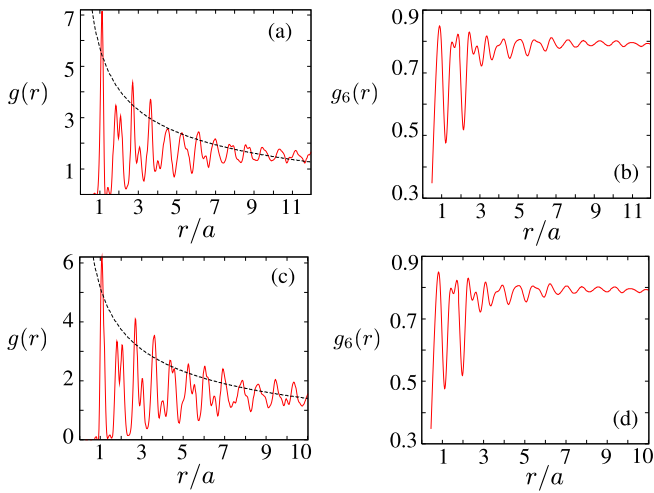}
\caption{\label{fig:correlation_functions} (color online)
Translational and
orientational correlation functions ($g$ and $g_6$, respectively) for
rafts with (a,b) $\kappa \approx 0.32$ cm$^{-1}$, $a=(0.8410 \pm 0.0025)$ mm,
and (c,d) $\kappa \approx 0.15$ cm$^{-1}$, $a=(0.9071 \pm 0.0037)$ mm.
All the curves are plotted as functions of $r/a$, where $r$ is
the planar distance from the center and $a$ is the bubble radius.
The envelope
for the crystalline solid decays algebraically (dashed line), while the orientational
correlation function approaches the constant value 0.8.}
\end{figure}

To characterize the order of the crystalline raft, we measure the
translational and orientational correlation functions $g(r)$ and
$g_{6}(r)$. The former gives the probability of finding a particle
at distance $r$ from a second particle located at the origin. The
function is normalized with the density of an equivalent homogeneous
system in order to ensure $g(r)=1$ for a system with no structure.
Interactions between particles build up correlations in their
position and $g(r)$ exhibits decaying oscillations, asymptotically
approaching one. For a two-dimensional solid with a triangular
lattice structure the radial correlation function is expected to
exhibit sharp peaks in correspondence with the sequence
$r/a=\sqrt{n^{2}+nm+m^{2}}=1,\,\sqrt{3},\,2,\,2\sqrt{3}\ldots$ while
the amplitude of the peaks decays algebraically as $r^{-\eta}$ with
$\eta=1/3$ \cite{ChaikinLubensky:2000} (dashed line in Fig.~\ref{fig:correlation_functions}). Within the precision of our data,
the positional order of the paraboloidal crystals assembled with the
bubble raft model reflects this behavior, although more accurate
measurements are required in order to clarify a possible dependence
of the exponent $\eta$ on the curvature.

The orientational correlation function $g_{6}(r)$ is calculated as
the average of the product $\langle \psi(0)\psi^{*}(r) \rangle$ of
the hexatic order parameter over the whole sample. For each bubble
(labeled $j$) that has two or more neighbors,
$\psi_{j}(r)=(1/Z_{j})\sum_{k=1}^{Z_{j}}\exp(6i\theta_{jk})$, where
$Z_{j}$ is the number neighbors of $i$ and $\theta_{jk}$ is the
angle between the $j-k$ bonds and a reference axis. One expects
$g_{6}(r)$ to decay exponentially in a disordered phase,
algebraically in a hexatic phase and to approach a non zero value in
the case of a crystalline solid. In the systems studied we find that
$g_{6}(r)$ to approaches value $0.8$ in the distance of $5$--$6$
lattice spacings.

Of particular interest is the structure of the grain boundaries
appearing in the paraboloidal lattice for different values of the
curvature parameter $\kappa$. Grain boundaries form in the bubble
array  during the growing process as a consequence of geometrical
frustration. As noted, any triangular lattice confined in a simply
connected region with the topology of the disk is required to have a
net disclination charge $Q=6$. Each disclination has an energy associated
with long-range elastic distortion of the lattice and a short range
core-energy. While the former is responsible for the emergent
symmetry of a geometrically frustrated crystal, the latter plays the
role of the energy-penalty required for the creation of a single
disclination defect. Although dependent on the interparticle
interactions, and so different from system to system, the
disclination core-energy is generally much smaller than the elastic
energy, especially in the case of a bubble array where the
particle-particle interaction is dominated by hard-core repulsion.
Defects are then favored to proliferate throughout the crystal. In
the flat plane, however, the elastic stress due to an isolated
disclination is extremely high and defects are energetically favored
to cluster in the form of a grain boundary consisting of
one-dimensional arrays of tightly bound $(5,7)-$fold disclinations
pairs. In a planar confined system, grain boundaries typically span
the entire length of the crystal, but if a non-zero Gaussian
curvature is added to the medium, they can appear in the form of
scars carrying a net $+1$ topological charge and terminating in the
bulk of the crystal.

\begin{figure}[b]
\centering
\includegraphics[width=1\columnwidth]{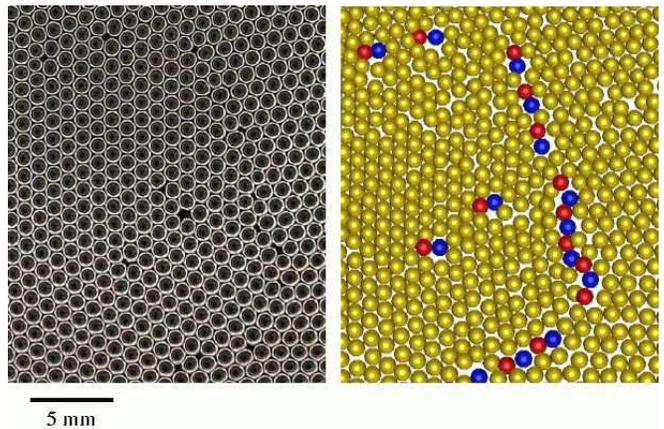}
\caption{\label{fig:scars} (color online)
Example of terminating grain boundary scar
for a system with large Gaussian curvature. The scar starts from the
circular perimeter of the vessel and terminates roughly in the center
carrying a net $+1$ topological charge.}
\end{figure}

Prominent examples of grain boundaries are visible in the two
lattices shown in Fig.~\ref{fig:raft}. For a gently curved
paraboloid (with $\kappa\approx 0.15$ cm$^{-1}$), grain boundaries
form long (possibly branched) chains running from one side
to the other and passing through the
center. As the curvature of the paraboloid is increased, however,
this long grain boundary is observed to terminate in the center
(see Fig.~\ref{fig:scars}). For $R=5$ cm,
the elastic theory of defects predicts
a structural transition at $\kappa_{c}=0.27$ cm$^{-1}$  in the limit
of large core energies \cite{GiomiBowick:2007}. In this limit the creation
of defects is strongly penalized and the lattice has the minimum number
of disclinations required by the topology of the embedding surface.
In a low curvature paraboloid ($\kappa<\kappa_{c}$)
these disclinations are preferentially located
along the boundary to reduce the elastic energy of the system. When
the aspect ratio of the paraboloid exceeds a critical value
$\kappa_{c}(R)$, however, the curvature at the origin is enough to
support the existence of a $5-$fold disclination and the system
undergoes a structural transition. In the limit of large core
energies, when only six disclinations are present, such a transition
implies a change from the $C_{6v}$ to the $C_{5v}$ rotational
symmetry group. Together with our experimental observations, these
considerations point to the following mechanism for scar
nucleation in a paraboloidal crystal. In the regime in which the
creation of defects is energetically inexpensive, geometrical
frustration due to the confinement of the lattice in a simply
connected region is responsible for the formation of a long
side-to-side grain boundary. But when the curvature of the paraboloid
exceeds a critical value (dependent on the radius of the circular
boundary), the existence of a $+1$ disclination near
the center is energetically favored. Such a disclination
serves as a nucleation site for $5-7$ dislocations and the
side-to-side grain boundary is replaced by a terminating
center-to-side scar. For intermediate curvature the theory also
predicts a regime of coexistence of isolated disclinations and
scars due to the variable Gaussian curvature. For dense systems
(i.e. number of subunits larger than a few hundred for our geometry),
the coexistence is suppressed because the embedding surface
will appear nearly flat at the length scale of a lattice spacing.
The bulk of the system is thus populated uniquely by scars. Our data
confirm this prediction.

\begin{figure}[t]
\centering
\includegraphics[width=1\columnwidth, clip]{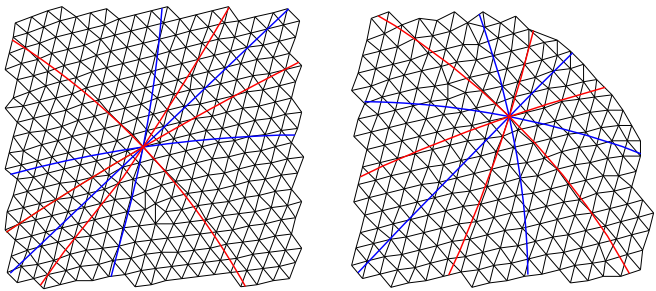}
\caption{\label{fig:geodesics} (color online) Delaunay
triangulation of a portion of the paraboloidal lattice with
$\kappa\approx 0.32$ cm$^{-1}$ near the center (left) and the 
boundary (right). Red and blue lines represent the geodesics
directed toward first and second neighbors, respectively.}
\end{figure}

Away from the center of the paraboloid, we have compared the
crystalline directions with the geodesics starting from a given
reference point (see Fig.~\ref{fig:geodesics}). Near the boundary,
the directions of both first and second neighbors (in red and blue
respectively), are reasonably aligned with the geodesics. The
alignment becomes decorrelated after roughly five lattice spacings
with the decorrelation more pronounced in the radial direction
(maximal principal curvature) where the normal curvature is largest.
As one gets closer to the center, the geodesic correlation becomes
weaker and almost completely vanishes along the radial direction.
Along the angular direction (minimal principal curvature), on the
other hand, the crystalline axes appear aligned with the geodesic
directions.

In this paper we have demonstrated a visualizable example of a
non-Euclidean, non-spherical crystal. Despite the simplicity of our
technique we found good agreement with the elastic theory of defects
in curved space. Bubble rafts are shown to be effective models for
the study of non-Euclidean crystallography. Bubble assemblages
provide a large number of particles (order $10^{3}$) with very 
simple and inexpensive equipment. They give access to details 
that are necessarily unavailable to continuous field theories 
and provide system sizes that are prohibitive for numerical 
simulations.

Future experiments might explore varying the shape of the vessel to
investigate the role of the boundary on the bulk order.  This setup 
is also suitable for studying dynamical phenomena such as the glide of 
dislocations in the relaxation process as well as the formation of 
vacancies and interstitials (e.g., following Ritacco \emph{et al.}
\cite{Ritacco:2007}, looking at a cascade of bursting
bubbles on a paraboloid).

We acknowledge David Nelson for stimulating discussions and Philip 
Arnold of the Syracuse University Physics Department machine shop 
for assisting in the construction of the 
experimental apparatus. The work of MJB, LG and HS was supported by 
the NSF through grants DMR-0219292 and DMR-0305407, and through funds 
provided by Syracuse University. The work of CKT was
supported in part by NSF grant DMR 0606424.


\begin{thebibliography}{99}

\bibitem{BraggNye:1947}
W. L. Bragg and J. F. Nye,
Proc. Roy. Soc. A \textbf{190}, 474 (1947).

\bibitem{Kittel:1966}
C. Kittel,
\emph{Introduction to solid state physics},
(Wiley, New York, 1966).

\bibitem{Feynman:1963}
R. P. Feynman, R. B. Leighton and M. Sands, 
\emph{Feynman lectures on physics Vol. 2}, 
(Addison-Wesley, Reading, Ma., 1963) Chapter 30.

\bibitem{SimpsonHodkinson:1972}
A. W. Simpson and P. H. Hodkinson,
Nature \textbf{237}, 320 (1972).

\bibitem{GouldstoneEtAl:2001}
A. Gouldstone, K. J. Van Vliet and S. Suresh,
Nature \textbf{411}, 656 (2001).

\bibitem{WangKrishanDennin:2006}
Y. Wang, K. Krishan and M. Dennin,
Phys. Rev. E \textbf{73}, 031401 (2006).

\bibitem{BowickNelsonTravesset:2000}
M. J. Bowick, D. R. Nelson and A. Travesset,
Phys. Rev. B \textbf{62}, 8738 (2000);
%
M. Bowick, A. Cacciuto, D. R. Nelson, and A. Travesset,
Phys. Rev. Lett. \textbf{89}, 185502 (2002);
%
A. Perez-Garrido, M. J. W. Dodgson, and M. A. Moore,
Phys. Rev. B \textbf{56}, 3640 (1997).

\bibitem{BowickEtAl:2004}
M. Bowick, D. R. Nelson, and A. Travesset,
Phys. Rev. E \textbf{69}, 041102 (2004).

\bibitem{VitelliNelson:2004}
V. Vitelli and D. R. Nelson,
Phys. Rev. E \textbf{70}, 051105 (2004).

\bibitem{VitelliLucksNelson:2006}
V. Vitelli, J. B. Lucks, and D. R. Nelson,
Proc. Natl. Acad. Sci. USA \textbf{103}, 12323 (2006).

\bibitem{GiomiBowick:2007}
L. Giomi and M. J. Bowick,
Phys. Rev. B \textbf{76}, 054106 (2007).

\bibitem{Diskpapers}
A. A. Koulakov and B. I. Shklovskii,
Phys. Rev. B \textbf{57}, 2352 (1998);
M. Kong, B. Partoens, A. Matulis and F. M. Peeters,
Phys. Rev E \textbf{69}, 036412 (2004);
A. Mughal and M. A. Moore,
Phys. Rev E \textbf{76}, 011606 (2007).

\bibitem{BauschEtAl:2003}
A. R. Bausch \emph{et al.}, 
Science \textbf{299}, 1716 (2003).

\bibitem{EinertEtAl:2005}
T. Einert \emph{et al.},
Langmuir \textbf{21}, 12076 (2005).

\bibitem{CrockerGrier:1996}
J. C. Crocker and D. G. Grier,
J. Colloid Interface Sci. \textbf{179}, 298 (1996).

\bibitem{ChaikinLubensky:2000}
P. M. Chaikin and T. C. Lubensky,
\emph{Principles of Condensed Matter Physics},
(Cambridge University Press, Cambridge, 2000).

\bibitem{Ritacco:2007}
H. Ritacco, F. Kiefer and D Langevin,
Phys. Rev. Lett. \textbf{98}, 244501 (2007).

\end{thebibliography}
\end{document}